\documentclass[12pt,preprint]{aastex}

\def\pc{{\rm pc}}
\def\kpc{{\rm kpc}}
\def\sech{{\rm sech}}

\def\brk{{\rm brk}}

\begin{document}

\title{Stellar Contribution to the Galactic Bulge Microlensing
Optical Depth}

\author 
{Cheongho Han}
\affil{Department of Physics, Institute for Basic Science Research,
Chungbuk National University, Chongju 361-763, Korea}
\email{cheongho@astroph.chungbuk.ac.kr}
\and
\author 
{Andrew Gould}
\affil{Department of Astronomy, Ohio State University, Columbus, OH
43210, USA}
\email{gould@astronomy.ohio-state.edu}

\singlespace

\begin{abstract}

We estimate the optical depth to self-lensing by stars in the Galactic
bulge using the {\it Hubble Space Telescope} star counts of 
Holtzman et al.\ and Zoccali et al.\ as extrapolated by Gould into the 
brown-dwarf and remnant regimes and deprojected along the 
line of sight using the model of Dwek et al.  We find a self-lensing
optical depth $\tau{\rm (bulge-bulge)}=0.98\times 10^{-6}$.  When
combined with the lensing of bulge stars by foreground stars in the
disk, this yields $\tau{\rm (bulge-total)}=1.63\times 10^{-6}$, in
reasonable agreement with the estimates of $\tau=2.13\pm 0.40\times 10^{-6}$ 
and $\tau=1.08\pm0.30\times 10^{-6}$ based on observations of clump giants
by the MACHO and EROS collaborations.

\end{abstract}
\clearpage
 
\section{Introduction
\label{sec:intro}}

The optical depth to microlensing toward the Galactic bulge has been
controversial since before it was first ``officially'' measured.
When \citet{pac91} and \citet{griest91}  first proposed bulge microlensing
observations, they estimated the optical depth to be about 
$\tau\sim 5\times 10^{-7}$ assuming all events were due to known disk stars.
However, as soon as the first six bulge events were reported by OGLE
\citep{ogle1,ogle2}, the apparently high event rate prompted 
\citet{kp} to evaluate the contribution to the optical depth of bulge stars 
in addition
to those of the disk.  They estimated $\tau\sim 8.5\times 10^{-7}$ and
concluded that the value could be as much as twice as high if the bulge
were elongated along the line of sight.

Nevertheless, the first measurements of the optical depth,
$\tau=3.3\pm 1.2 \times 10^{-6}$ by OGLE \citep{ogle3}, and
$\tau=3.9^{+1.8}_{-1.2} \times 10^{-6}$ by MACHO
\citep{macho1}\footnote{Submission
of the manuscript was delayed 4 months by a secretarial
error (C.\ Alcock 1996, private communication) but the results were
widely circulated in the microlensing community (astro-ph/9512146).
} were substantially higher than even the highest predictions.
Indeed, \citet{gould94} and \citet{kk} argued that they were so high
as to eliminate the need for any dark halo inside the solar circle, 
provided the Galaxy was assumed axisymmetric.  This argument was subsequently
generalized to non-axisymmetric mass distributions by \citet{binney}.

These early results motivated substantial additional work.  On the one
hand MACHO made several new estimates of the bulge optical depth,
one based on a difference-image analysis of a large number of primarily
faint sources, yielding $\tau=3.2\pm 0.5\times 10^{-6}$ 
\citep{macho2}, and another
based on clump giants, yielding $\tau=2.0\pm 0.4 \times 10^{-6}$ 
\citep{macho3}.  On the
other hand, beginning with \citet{zsr}, many workers attempted to develop 
highly non-axisymmetric bulge models that could account for the high 
observed optical depth \citep{metcalf,zrs,zm,bebg,gyuk,nm,sk,be}.

Most recently, EROS has reported a much lower optical depth,
$\tau=0.94\pm 0.30\times 10^{-6}$ 
\citep{eros} based on a study specifically designed to
monitor clump giant sources as advocated by \citet{gould95}.  In their
Table 2, \citet{eros} also put all optical-depth measurements
on a common basis by adjusting them for their offset from Baade's Window (BW).
Hence, for example, their own measurement was adjusted upward to
$\tau=1.08\pm 0.30\times 10^{-6}$ because their mean field position was
less densely populated than BW.

Given the strongly divergent observational results as well as the
theoretical difficulties these pose, it is useful to have as many
model-independent constraints as possible.  One such constraint
comes from star counts.  At the time that \citet{pac91}, \citet{griest91},
and \citet{kp} made their estimates, the stellar content of the disk
was poorly measured and the stellar content of the bulge was almost
completely unconstrained.  Star count from {\it Hubble Space Telescope (HST)}
observations have now completely transformed that situation
\citep{holtzman,zoccali,zheng}.  The stellar mass function is
well measured  in the disk down to the hydrogen-burning limit and in the
bulge down to $M\sim 0.15\,M_\odot$.  There are, of course, uncertainties
in extrapolating these mass functions to lower, sub-stellar masses, and to
the remnants of the now deceased stars at higher masses.  Nevertheless,
the fact that large portions of these mass functions have been directly
measured allows one to construct useful constraints.

We formulate these constraints within the context of a specific
model of the Galactic bulge and disk.
The model of the disk is relatively secure, and uncertainties in it
play an overall very small role in controversies about the total 
optical depth toward the bulge.  That is why most of the effort to explain 
the high optical depth has centered on models of the bulge.  The benchmark
model that we adopt is therefore far from unique.  We show, however, 
that it is
possible to factor the stellar constraint into two terms, one
representing the bulge model convolved with the observational strategy, 
and the other representing the star counts.  In this way, our result can
easily be applied to any model of the Galactic bulge.

\section{Bulge and Disk Models
\label{sec:models}}

\subsection{Disk
\label{sec:disk}}

	We model the local vertical disk density profile in accord with 
the model of \citet{zheng}.  To extend this model to the whole
Galactic disk, we assume that the column density of the disk has
a scale length $H=2.75\,\kpc$, as measured by \citet{zheng}.  We
account for the gradual flaring of the disk by rescaling the scale heights
in the \citet{zheng} formula in proportion to the scale height derived
by \citet{kent}.  We normalize the local stellar column density to
$\Sigma_0 = 36\,M_\odot\,\pc^{-2}$.  
This includes $28\,M_\odot\,\pc^{-2}$ in observable stars and white
dwarfs \citep{zheng,gbf} and another $8\,M_\odot\,\pc^{-2}$, which is
a rough estimate of the column density of brown dwarfs (BDs).  
The disk density profile in cylindrical coordinates is then,
\begin{equation}
\rho(R,z) = {\rho_0 \over\eta} \exp\biggl(-{R-R_0\over H}\biggr)
\biggl[(1-\beta)\sech^2{z\over \eta h_1} + 
\beta\exp\biggl(-{|z|\over \eta h_2}\biggr)\biggr]
\label{eqn:zhengprofile}
\end{equation}
where $\rho_0=0.0493\,M_\odot\,\pc^{-3}$, $\beta = 0.565$,
$h_1=270\,\pc$, $h_2=440\,\pc$, $H=2.75\,\kpc$, $R_0=8\,\kpc$, and
\begin{equation}
\label{eqn:}
\eta(R) = {\rm max}\biggl\{{R\over 9025\,\pc} + 0.114, 0.670\biggr\}.
\end{equation}

\subsection{Bulge
\label{sec:bulge}}

For the bulge model, we follow \citet{hg95} and scale the bulge mass
density to the deprojected infrared light density profile of
\citet{dwek}.  Specifically, we use model G2 (with $R_{\rm max}=5\,\kpc$)
from their Table 2.  However, whereas \citet{hg95} fixed the normalization
of this model using the tensor virial theorem (which is sensitive to
all the mass), we will normalize it by the observed stars (and inferred
stellar remnants and substellar objects).

\subsection{Mass Function and Mass-Luminosity Relation
\label{sec:mf}}

For the bulge, we adopt the (unnormalized) mass function of \citet{gould00}.
This assumes that bulge stars formed initially according to
a power law, $d N/dM = k(M/M_\brk)^\alpha$, where $M_\brk=0.7\,M_\odot$,
$\alpha=-2.0$ for $M>M_\brk$, and $\alpha=-1.3$ for $M<M_\brk$.
These slopes are consistent with the observations of \citet{zoccali},
but the profile is extended below their lower limit of $M\sim 0.15\,M_\odot$
to a BD cutoff of $M=0.03\,M_\odot$.  The stars with
initial masses $1\,M_\odot<M<8\,M_\odot$ are assumed to have become
white dwarfs (WDs), those with masses $8\,M_\odot<M<40\,M_\odot$ are 
assumed to have become neutron stars (NSs) with masses $M=1.35\,M_\odot$, 
and those with masses $M>40\,M_\odot$ are assumed to have become black
holes (BHs) with mass $M=5\,M_\odot$.  The total mass
is then dominated by main-sequence (MS) stars, with mass fractions
\begin{equation}
{\rm BD:MS:WD:NS:BH} = 7:62:22:6:3.  
\label{eqn:ratios}
\end{equation}
This is important because, as we will see in
\S~\ref{sec:normal},  the observational constraint comes from luminous 
(i.e. MS) stars.

We adopt the mass-$M_V$ relation of \citet{allen}

\section{Normalization From Star Counts
\label{sec:normal}}

\begin{figure}
\plotone{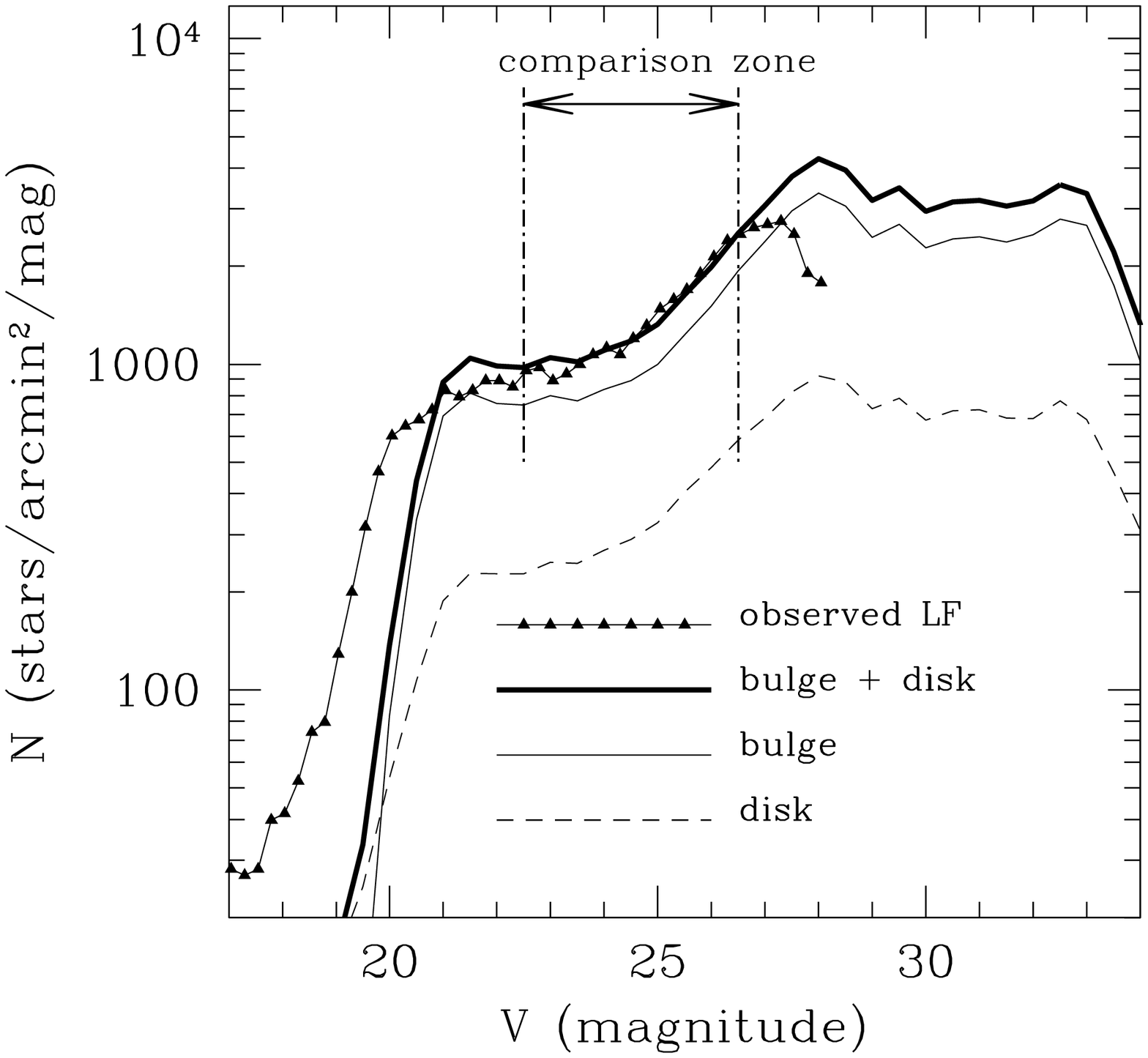}
\caption{\label{fig:lf}
Star counts toward Baade's Window.  Observations of \citet{holtzman}
({\it connected data points}) are used to fix the normalization
of the bulge mass-density profile of \citet{dwek}.  The individual 
contributions of the disk ({\it dashed curve}) and bulge ({\it solid curve})
are added to predict the total disk+bulge counts ({\it bold curve}).
The disk is held fixed, while the bulge is scaled until the mass-weighted
disk+bulge star-count predictions agree with the
observations over the range $22.5<V<26.5$.
}\end{figure}

We populate the bulge with stars (and BDs and
remnants) according to the \citet{gould00} mass function described in
\S~\ref{sec:mf} and adjust the overall normalization until we match
the \citet{holtzman} $V$-band star counts in BW.  See Figure \ref{fig:lf}.  
To predict these counts, we assign each star a luminosity using
the \citet{allen} mass-$M_V$ relation, while treating all BDs, WDs,
NSs, and BHs as dark.  We then convert to apparent magnitudes using
the each star's individual distance.  Finally, each star is then reddened 
by assuming that the total extinction along the line of sight is $A_V=1.28$
\citep{holtzman}, and that the dust has a scale height of 120 pc.

We incorporate disk as well as bulge stars in predicting the \cite{holtzman}
star counts.  Since the disk stellar profile is regarded as well
measured (by \citealt{zheng}), we do not adjust the normalization
of the disk profile as we do the bulge profile, but rather leave it
fixed in the form given in \S~\ref{sec:disk}.  However, we use the
same mass function and mass-$M_V$ relation for the disk as the bulge.
In principle, one should make an independent estimate of these functions.
However, since the disk stars contribute only $\sim 15\%$ of the counts
(see Fig.~\ref{fig:lf}), the net corrections from more accurate functions
would be only a few percent, which is small compared to other uncertainties
in the problem.  Hence, we ignore this distinction in the interest of
simplicity.

	We fix the normalization by demanding agreement between the
model predictions and the (mass-weighted) \citet{holtzman} star counts 
over the range $22.5<V<26.5$.  
At fainter magnitudes, the \citet{holtzman} data become seriously
incomplete, while at somewhat brighter magnitudes our simple mass-$M_V$
relation fails to account for evolution off the MS and so slightly
overpredicts the counts.  At much brighter magnitudes, the fact that
our model has no giants causes it to completely underestimate the
counts.  These latter two effects are each small and roughly cancel one
another.  See Figure~\ref{fig:lf}.

	The good agreement over four magnitudes demonstrates that
the mass function of \citet{zoccali}, which is based on infrared
observations, is compatible with the optically-based \citet{holtzman}
mass function, an agreement already noted by \citet{zoccali}.

	The total column density of bulge stars (and associated BDs and
remnants) toward BW is
\begin{equation}
\Sigma_* = 2086\,M_\odot\,\pc^{-2}.
\label{eqn:sigmastar}
\end{equation}
As shown by equation~(\ref{eqn:ratios}), 62\% of this mass is in the 
form of luminous stars, while 55\% is in MS stars with masses 
$0.15\,M_\odot<M<1\,M_\odot$,
and so with sufficient luminosity to have been directly observed by 
\citet{zoccali}.

\section{Optical Depth Due to Stars
\label{sec:optdep}}

The observed optical depth will always be an average over the individual
optical depths to the stars being monitored.  More distant stars have
higher optical depth, but are also fainter and so less likely to
be included in the sample \citep{kp}.  To parameterize this effect, we
write
\begin{equation}
\langle \tau\rangle_\gamma = {4\pi G\over c^2}{
\int_0^\infty d D_s D_s^{2-\gamma} \rho(D_s)
\int_0^{D_s}d D_l \rho(D_l) D_l (D_s - D_l)/D_s
\over
\int_0^\infty d D_s D_s^{2-\gamma} \rho(D_s)
}
\label{eqn:taugamma}
\end{equation}
For standard candles, which can be identified independent of distance,
$\gamma=0$.  We are most directly interested in comparing to optical
depth measurements using clump giants, which are approximately standard
candles.  We therefore adopt $\gamma=0$ as the default.  For bulge
sources and for, respectively, disk and bulge lenses, we find
\begin{equation}
\langle \tau\rangle_0({\rm disk}) = 0.65\times 10^{-6},\qquad
\langle \tau\rangle_0({\rm bulge}) = 0.98\times 10^{-6}.
\label{eqn:tau0}
\end{equation}
In fact, $\langle \tau\rangle_\gamma({\rm disk})$ does not significantly
depend on $\gamma$, but $\langle \tau\rangle_\gamma({\rm bulge})$ does.
For example $\langle \tau\rangle_1({\rm bulge})=0.86\times 10^{-6}$

The total optical depth due to stars that is predicted by this mass
profile for observations toward BW, $\langle \tau\rangle_0({\rm total}) = 
1.63\times 10^{-6}$, agrees reasonably well with the two measurements
made using clump giants.  When these are adjusted to a common mean
direction at BW \citep{eros}, they yield 
$\tau=2.13\pm 0.40\times 10^{-6}$ \citep{macho3} and
$\tau=1.08\pm 0.30\times 10^{-6}$ \citep{eros}. 

\section{Discussion
\label{sec:discuss}}

While the agreement between the observations and the model predictions
is comforting, it is important to recognize that the model has some
uncertainties.  These are most easily discussed by writing the predicted
optical depth as
\begin{equation}
\langle \tau\rangle_\gamma({\rm bulge}) 
= {4\pi G\Sigma_* \overline{D_\gamma}\over c^2}.
\label{eqn:dbar}
\end{equation}
This equation serves to define $\overline{D_\gamma}$, which may be
thought of roughly as the characteristic source-lens separation.
This characteristic separation depends only on the mass profile and
(through $\gamma$) on the observational strategy.  It can be
evaluated trivially for any mass model.  For the \citet{dwek} bulge
model used here,
\begin{equation}
\overline{D_0} = 782\,\pc.
\label{eqn:d0eval}
\end{equation}

	On the other hand,
$\Sigma_*$ depends only on the surface density of stars.  
For luminous stars, this is practically an observed quantity
\citep{holtzman,zoccali}, and the major uncertainty is how
to extrapolate the observations into the BD and remnant regimes.
One might, for example, argue that a Salpeter mass function $(\alpha=2.35)$
is more appropriate for the extrapolation to higher masses. In this
case, the WD contribution would fall by 20\% and the BH contribution
would fall by a factor 5.  Or one could argue that the mean mass of
BHs is higher, or that the threshold stellar mass to leave BH remnants
is lower than the $40\,M_\odot$ adopted here, either of which would raise 
the BH contribution.  Finally, the \citet{zoccali} mass function does
not take account of binary companions, which might plausibly raise $\Sigma_*$
by 10--20\%.  Any of these changes can be easily incorporated 
using equation~(\ref{eqn:ratios}), once new estimates are made.

We can compare the approach adopted here to that of \citet{hg95}, who
normalized the \citet{dwek} model using the tensor virial theorem.
To do so, we first note that \citet{hg95} tilted their bulge ellipsoid
$20^\circ$ to the line of sight, as advocated in the \citet{dwek} abstract,
rather than the $13.\hskip-2pt ^\circ 4$ given in Table 1 of 
\citet{dwek} and used here.  Normalizing the \citet{hg95} $20^\circ$ model 
according to the procedure defined in this paper, we find a bulge
self-lensing optical depth of $\langle\tau\rangle_0=0.88\times 10^{-6}$,
only 66\% of the value obtained using the \citet{hg95} tensor virial theorem
normalization. Part of
the difference is undoubtedly due to binary companions, which contribute to the
integrated mass and enter the
microlensing optical depth but which, because they are generally substantially
less luminous than their primaries, leave almost no trace on the observed
{\it HST} stellar luminosity functions.  If the BD or remnant populations
were substantially larger than modeled by \citet{gould00}, this would
also add to both the integrated mass density and the optical depth.
Part of the difference could also be due to non-baryonic dark matter, which
would add to the integrated mass (to which the tensor virial theorem is
sensitive) but would not contribute to microlensing.  Finally, of course,
the tensor virial theorem yields results that are only as accurate as
the mass model and velocity dispersion tensor to which it is applied,
and either of these could have errors.


\acknowledgments 
Work by CH was supported by the Astrophysical Research Center
for the Structure and Evolution of the Cosmos (ARCSEC) of Korea
Science and Engineering Foundation (KOSEF) through the Science
Research Center (SRC) program.
Work by AG was supported by grant AST 02-01266 from the NSF.

\clearpage

\end{document}